\documentclass[preprint]{aastex}

\begin{document}

\title{Subaru Deep Spectroscopy of a Star-forming Companion Galaxy of BR
1202-0725 at $z=4.7$
\footnote{Based on data collected at the Subaru Telescope, which is
operated by the National Astronomical Observatory of Japan.}
}

\author{Youichi Ohyama\altaffilmark{2}, Yoshiaki Taniguchi\altaffilmark{3}, \&
Yasuhiro Shioya\altaffilmark{3}}

\altaffiltext{2}{Subaru Telescope, National Astronomical Observatory
            of Japan, 650 N. A`ohoku Place, Hilo, HI 96720}

\altaffiltext{3}{Astronomical Institute, Graduate School of Science,
           Tohoku University, Aramaki, Aoba, Sendai 980-8578, Japan}

\begin{abstract}
We present deep spatially-resolved optical spectroscopy of the NW companion
galaxy of the quasar BR 1202-0725 at $z=4.7$.
Its rest-frame UV spectrum shows star-forming activity in the nuclear region.
The Ly$\alpha$ emission profile is symmetric with wavelength
while its line width is unusually wide (FWHM $\simeq 1100$ km s$^{-1}$) for
such a high-$z$ star-forming galaxy.
Spectrum taken along the Ly$\alpha$ nebula elongation, which is almost along
the minor axis of the companion host galaxy, reveals that off-nuclear
Ly$\alpha$ nebulae have either flat-topped or multi-peaked profiles along
the extension. All these properties can be understood in terms of 
superwind activity in the companion galaxy.
We also find a diffuse continuum component around the companion, which shows
similar morphology to that of Ly$\alpha$ nebula, and is most likely due to
scattering of the quasar light at dusty halo around the companion.
We argue that the superwind could expel dusty material out to the halo region,
making a dusty halo for scattering.
\end{abstract}

\keywords{quasars: individual (BR 1202-0725) --- galaxies: starburst}

\section{INTRODUCTION}

BR 1202$-$0725 at $z=4.7$ is one of well-studied quasars at high redshift
(e.g., Kennefick, Djorgovski, \& Meylan 1996; Ohta et al. 1996;
Omont et al. 1996; Hu, McMahon, \& Egami 1996; Petitjean et al. 1996;
Storrie-Lombardi et al. 1996).
A faint companion galaxy is also found to be associated with BR 1202-0725
(Djorgovski 1995).
This is located at 2.3\arcsec~ NW of the quasar (hereafter NW companion)
and shows an extended Ly$\alpha$ nebula
almost at the same redshift as that of the quasar (Hu et al. 1996;
Petitjean et al. 1996; Storrie-Lombardi et al. 1996).
Previous optical spectroscopies show that it is a star-forming galaxy with
only a narrow Ly$\alpha$ emission (no strong metal emission lines, such as
N {\sc v}, C {\sc iv}) (e.g., Petitjean et al. 1996;
Hu, McMahon, \& Egami 1997; Fontana et al. 1998).
The redshifted [O {\sc ii}] emission is also detected on the galaxy
at near-infrared wavelength (Ohta et al. 2000; see also Pahre \& Djorgovski
1995 for earlier measurement with only an upper-limit value).
Interestingly, redshifted CO emission is detected at another location near
the quasar (4\arcsec~ NW of the quasar), being $\simeq 2$\arcsec~ away from
the NW companion (hereafter, 2nd CO emitter), as well as on the quasar itself
(Omont et al. 1996; Carilli et al. 2002; see also Ohta et al. 1996).
There are lines of evidence for vigorous star-forming activities at quasar and
2nd CO emitter (e.g., massive content of molecular gas, high excitation
temperature of CO lines, optical-FIR SED which is typical of vigorous
star-forming galaxies) (e.g., Benford et al. 1996; Omont et al. 1996;
Ohta et al. 1996; Ohta et al. 1998; Ohta et al. 2000; Yun et al. 2000;
Carilli et al. 2002).
Therefore, the ``BR 1202-0725 group'' seems to contain various kinds of
young star-forming objects, and is suitable for investigating star-forming
activities in early universe.

Among the three objects in the BR 1202-0725 group, the NW companion would give
us an unique opportunity to investigate nature of star formation in a young
galaxy
because this galaxy appears to be free both from extremely
dusty environment and from intense quasar light.
Therefore we have made very deep optical spectroscopy of the NW
companion with the 8.2m Subaru Telescope, and present our new results
in this paper.
We assume
$\Omega_{\rm M}=0.3, \Omega_\Lambda=0.7, H_0=70$ km s$^{-1}$ Mpc$^{-1}$
in this paper.
1\arcsec~corresponds to 6.7 kpc, in the adopted cosmology.

\section{OBSERVATION AND DATA REDUCTION}

We used the FOCAS (Kashikawa et al. 2002) attached at Cassegrain focus of
the Subaru Telescope (Iye et al. 2004) on Feb. 14 and 15, 2004.
A VPH grism (with 600 grooves mm$^{-1}$ and 6500\AA~central wavelength) and
an order-sorting filter Y47 were used to cover a wavelength range from
5950\AA~to 8400\AA.
With a combination of a 0.8\arcsec~width longslit, this setting provides a
spectral resolution of $R=1700$ (measured with sky lines) near the redshifted
Ly$\alpha$ ($\simeq 6930$\AA).
The CCDs were binned onchip to $3\times 2$ (0.3\arcsec~$\times$ 0.2\arcsec~in
space and wavelength directions, respectively).
The slit was placed on the companion at two position angles (PAs), at
PA$=-38.08^{\circ}$ (along the quasar and the NW companion: Hu et al. 1996;
hereafter, NW-SE slit) on the first night, and at PA$=-128.08^{\circ}$
(perpendicular to the NW-SE slit; hereafter, NE-SW slit) on the second night
(Figure 1).
We took eight 30 minute exposures per slit PA (four hours in total).
A small nodding of the target along the slit was applied during each exposure.
Seeing size was $\simeq 0.5$\arcsec$-0.8$\arcsec~FWHM during the observations,
and was typically 0.6\arcsec~FWHM.

Data reduction was made in a standard manner, i.e., bias subtraction, flat
fielding, wavelength calibration with Th-Ar arc lines, sky subtraction, and
spectral sensitivity calibration with a standard star GD153, were applied.
Atmospheric absorption features were corrected with another spectrum of GD153,
taken with exactly the same spectroscopy setting as that for the companion.
Then, a one-dimensional nuclear spectrum was extracted from the two-dimensional
spectra in the following way.
First, two nuclear spectra taken at each slit PA were extracted over
0.6\arcsec~aperture along the slit around the peak, and then two spectra were
coadded.
Then, a contamination of the bright quasar light is corrected, by subtracting
the scaled quasar spectrum from the observed spectrum.
Here the scale of the quasar spectrum was estimated by measuring the quasar
flux at the same distance to the companion but at another side of the quasar
(2.3\arcsec~SE of the quasar) on the spectrum taken along the NW-SE slit.
Figure 2 shows the final companion nuclear spectrum as well as the quasar one.
Further, we deduced one-dimensional spatial flux distribution of both
continuum and Ly$\alpha$.
For each slit PA, two-dimensional spectrum is averaged over the wavelength
range of $\lambda > \lambda_{\rm Ly\alpha}$ and just around
$\lambda_{\rm Ly\alpha}$ to obtain spatial flux distributions of continuum
and Ly$\alpha$, respectively (Figure 3).

\section{RESULTS}

The nuclear spectrum (Figure 2) is composed of a narrow Ly$\alpha$ emission,
almost flat continuum emission at $\lambda > \lambda_{\rm Ly\alpha}$, and
partially absorbed continuum at $\lambda < \lambda_{\rm Ly\alpha}$.
The Ly$\alpha$ emission is detected at $\lambda_{\rm peak}=6932$\AA~(or
$z=4.7026$), and its width is $\simeq 1100$ km s$^{-1}$ FWHM, being
consistent with previous results (Petitjean et al. 1996; Fontana et al. 1998).
The line profile is almost symmetric along wavelength, except for a narrow
absorption line at blue side of the profile ($z=4.687$, or
$\Delta V\equiv V-V_{\rm Ly\alpha~peak}\simeq -800$ km s$^{-1}$), and can
indeed be reproduced with a Gaussian emission affected by a single absorption
line (Figure 2).
We point out that, although Petitjean et al. (1996) showed the blue-deficient
asymmetric profile of Ly$\alpha$, such profile seems to be a result of
unresolved blue absorption in their lower-quality spectrum.

The nuclear Ly$\alpha$ flux is $f$(Ly$\alpha$ nuc.)$=6.5 \times 10^{-17}$
erg s$^{-1}$ cm$^{-2}$, which is significantly smaller than that of the total
one ($f$(Ly$\alpha$ total)$=2.7 \times 10^{-16}$ erg s$^{-1}$ cm$^{-2}$).
Note that the total flux, rather than the nuclear one, is consistent with
the previously reported values (Hu et al. 1996; Petitjean et al. 1996;
Fontana et al. 1998).
The continuum spectrum at red side of Ly$\alpha$
($\lambda > \lambda_{\rm Ly\alpha}$) shows neither absorption nor emission
lines within the observed wavelength range, although rather poor S/N of
the continuum spectrum hampered detection of rather week absorption lines.
The continuum flux at 1400\AA, without extinction correction, is
$f(\rm continuum)=7.3 \times 10^{-32}$ erg s$^{-1}$ cm$^{-2}$ Hz$^{-1}$.
At blue side of Ly$\alpha$ ($\lambda < \lambda_{\rm Ly\alpha}$), the spectrum
shows absorbed continuum, which looks very similar to the quasar one showing
rich Ly$\alpha$ absorptions and a DLA system (e.g.,
Storrie-Lombardi et al. 1996).

We found that both the Ly$\alpha$ and the continuum emissions come from
extended region out to $3$\arcsec$ - 4$\arcsec~from their peaks along the
NE-SW slit, and their flux peaks coincide with each other (Figure 3).
This result is slightly different from that of Hu et al. (1996), where they
showed that the peak of the Ly$\alpha$ nebula is located about 0.6\arcsec~E
from the continuum peak.
However, the nebula elongation (NE) appears to be closer to the direction of
the peak displacement (E) mentioned by Hu et al. (1996).
The Ly$\alpha$ nebula seems to show more centrally concentrated flux
distribution comparing with that of the continuum along the slit,
especially at the SW side of the companion.
The NE nebula is brighter and more extended from the nucleus (distance from
the nucleus: $r\simeq 4$\arcsec~or $\simeq 27$ kpc).
At $r= 0.5$\arcsec$-1.5$\arcsec~NE, the emission shows flat-topped profile,
which is remarkably different from one at nucleus (Figure 4).
We found that the profile can be reproduced by a combination of two Gaussian
components (blue and red) at $\Delta V \simeq \pm 400$ km s$^{-1}$, and the
line width of each component is $600-1200$ km s$^{-1}$ FWHM.
At even outer NE regions ($r=2$\arcsec$-3.5$\arcsec~NE), the emission shows
more complicated shapes, and the profile is probably composed of three velocity
components (a near-systemic one at $\Delta V \simeq +200$ km s$^{-1}$, a blue
one at $\Delta V \simeq -700$ km s$^{-1}$, and a red one at
$\Delta V \simeq +500$ km s$^{-1}$) (see spectrograms in Figure 1, as well as
Ly$\alpha$ line profiles in Figure 4).
The overall line width, including these three components, is as wide as
$\sim 1500$ km s$^{-1}$, and the mean velocity is close to the systemic
velocity ($|\Delta V| \lesssim 200$ km s$^{-1}$).
At another side of the companion, the SW nebula, is fainter and more compact
in space ($\simeq 3$\arcsec~or $\simeq 20$ kpc).
At $r=0.5$\arcsec$-2.5$\arcsec~SW, the profile looks similar to that at outer
NE nebula, although the profile looks composed of two brighter components at
blue ($\Delta V\simeq -400$ km s$^{-1}$) and near-systemic velocity
($\Delta V \simeq 0$ km s$^{-1}$) as well as a possible fainter component at
red ($\Delta V \sim +500$ km s$^{-1}$) which is barely visible on a smoothed
spectrogram.

Compared with the NE-SW extension, the extension along the NW-SE slit is much
less prominent, although Ly$\alpha$ nebula is slightly more extended than that
for the continuum.
There is an extension toward NW (toward 2nd CO emitter) out to
$r \simeq 2$\arcsec~($\sim 14$ kpc).
The profile there looks more complicated than that the NE nebula,
and is probably composed of two narrow components with overall velocity extent
of as wide as 1000 km s$^{-1}$ at its mean velocity of
$\Delta V \sim +500$ km s$^{-1}$.
At another side of the quasar toward SE, fainter and narrow ($\sim 500$
km s$^{-1}$ FWHM) extension is detected ($r \simeq 1$\arcsec, or $\simeq 7$
kpc).
There is no major velocity structure along the extension
($|\Delta V| \lesssim 200$ km s$^{-1}$), although details of its kinematical
properties are not well known due to contamination of brighter quasar light.
These properties (extent and velocity structure) of the nebula along the
NW-SE slit are consistent with the report of Fontana et al. (1998).
We also note that there is a nebula extension even at SE of the quasar
($r \sim 4$\arcsec) at $\Delta V \sim -500$ km s$^{-1}$ (see a spectrogram
in Figure 1).
Since we do not have any detailed information for it, no discussions will
be made on it.

\section{DISCUSSION}

\subsection{Star-Formation Activity of the NW Companion}

The NW companion shows the following distinct characteristics:
(1) It has a narrow [FWHM(NW companion)$<<$FWHM(quasar)] Ly$\alpha$ emission,
whose redshift is very close to that of quasar.
(2) Its continuum spectrum shows a step in flux between blue and red sides of
Ly$\alpha$.
(3) It has no strong metal emission lines, such as N {\sc v} and C {\sc iv},
being suggestive of AGN activity.
All these indicate that the companion is a star-forming galaxy which is
physically associated with BR 1202-0725 quasar.
We estimate the star-forming rate (SFR) to be $\simeq 13$ $M_{\rm \sun}$
yr$^{-1}$ based on the relation of Kennicutt (1998) and the estimated
continuum luminosity at 1500\AA~from our measurement at 1400\AA.
Note that the Ly$\alpha$ luminosity seems not useful for estimating SFR,
because Ly$\alpha$ emission from the extended nebula, which is likely to be
excited by shock, may contribute to the nuclear Ly$\alpha$ flux (see later
sections for details).

Properties of the companion host galaxy is examined by using a stellar
continuum color of $I$ and $K$.
We do not use $R$ data, because it includes a strong Ly$\alpha$ emission and
bluer-than-Ly$\alpha$ continuum, both of which are difficult to be modeled.
We adopt $I=24.1\pm 0.2$ and $K=23.4\pm 0.4$ from Fontana et al. (1996) and
Hu et al. (1996), and a contribution of redshifted [O {\sc ii}]$\lambda$3727 in $K$
is corrected based on [O {\sc ii}] equivalentwidth measured by Ohta et al. (2000).
Continuum emission from the ionized gas is neglected here, because
such continuum emission is generally much fainter than that of stellar
continuum in starburst galaxies.
We adopt a synthetic stellar spectrum model ``starburst 99''
(Leitherer et al. 1999) to match the observed color.
We adopt instantaneous burst models (IMF slope $\alpha=2.35$,
$M_{\rm up}=100$ M$_{\rm \odot}$, $M_{\rm low}=1$ M$_{\rm \odot}$,
$Z=1/5Z_{\rm \sun}$ or $Z=1/20Z_{\rm \sun}$) with various ages (5-50 Myr).
Here we fix IMF parameters, and ignore the reddening effect for simplicity.
By comparing models and the observed color, we found that models of lower
metallicity ($\simeq 1/20 Z_{\rm \sun}$) and younger age (10 Myr) better
represent the observations (Figure 5).
It seems important to note that the observed blue UV color can only be
reproduced if almost no reddening is applied on the young stellar population
with bluest color.

\subsection{Nature of the Extended Ly$\alpha$ Nebula}

\subsubsection{Kinematic Properties of the Nebula}

Although the NW companion shows characteristic properties for star-forming
galaxies, it also shows unusual properties for such a high-$z$ star-forming
galaxy:
(1) At the nucleus, Ly$\alpha$ emission profile is wide ($\simeq 1100$
km s$^{-1}$ FWHM) and almost symmetric in wavelength, being different from
narrower ($200-400$ km s$^{-1}$ FWHM) and blue-deficient asymmetric profiles
which are typical for such a high-$z$ star-forming galaxy (e.g.,
Dawson et al. 2002; Ajiki et al. 2002; Kodaira et al. 2003; Rhoads et al. 2003;
see for a review Taniguchi et al. 2003 and references therein).
(2) At the off-nucleus, the companion show an extended Ly$\alpha$ nebula with
a complicated velocity structure, whose overall velocity width is as wide as
$\sim 1500$ km s$^{-1}$
\footnote[2]{
Since Ly$\alpha$ is a resonance line, the observed Ly$\alpha$ profile might be
strongly affected by the neutral gas around the companion (see, e.g.,
Mas-Hesse et al. 2003).
However, since nuclear Ly$\alpha$ profile does not show a blue-deficient
profile, and the overall velocity extent of the Ly$\alpha$ profile (e.g.,
blue/red tip velocity of the profile) is almost symmetric around the peak of
the nuclear Ly$\alpha$ profile at off-nuclear regions (except for NW and SW
tip of the nebula), we think that effect of the neutral gas for the observed
Ly$\alpha$ profile could be rather small.
Therefore, we assume in the following that the observed Ly$\alpha$ profile is
determined by kinematical and/or excitation structures of the ionized gas.
}.
If we assume that an extended nebula with such a violent kinematical status
surrounds the companion, we could expect that the high velocity component
($|\Delta V| > 300$ km s$^{-1}$) on the nucleus arises from the nebula along
our line-of-sight toward the companion nucleus.
If this is the case, the NW companion is likely composed of a usual
star-forming galaxy and the surrounding extended nebula with violent
kinematical status.

Here a question arises as what is the origin of the extended nebula.
There seems several possibilities:
(1) scattering of the quasar light by circumgalactic gas,
(2) photoionization of circumgalactic gas by the quasar UV light,
(3) spatially extended star-formation of the companion,
(4) cooling radiation from protogalaxies within dark matter halos (e.g.,
Haiman, Spaans, \& Quataert 2000), and
(5) galaxy-scale shock heating.
First possibility can be rejected due to different spectrum shape between
the nebula and the quasar.
Second possibility seems less likely since no strong emission lines suggestive
of AGN excitation, such as N {\sc v}, are detected.
Third and fourth possibilities seem also less likely since they could not
explain straightforwardly the observed kinematical structure of the nebula,
such as wide line width and flat-topped/multi-peaked profiles.
Also, the third possibility is less likely because the host galaxy elongates
along NW-SE directions (Hu et al. 1996), being perpendicular to the direction
of the nebula elongation.
Therefore, the last possibility seems most feasible, since shock could be
excited within the nebula showing the observed violent kinematical structure,
and it can emit intense Ly$\alpha$ emission (e.g., Shull \& McKee 1979),
if we assume that some kind of mechanism works to produce a violent internal
motion of the nebula.

We showed that the Ly$\alpha$ profile in the NE nebula can be composed of two
(blue and red) or three (blue, near-systemic, and red) components at inner and
outer parts of the nebula, respectively (see section 3).
We note that the inner nebula, which appears to show only two components,
could be composed of three components, and the fainter and narrower
near-systemic one was missed due to nearby brighter and wider components.
The SW nebula also seems to have two (blue and near-systemic) components,
and a possibly fainter red component.
One may think that these nebulae show similar kinematical properties, and
might form a single, large, elongated nebula at both sides of the companion
along its minor axis.
An idea to explain a pair of blue and red components at a same position
within the nebula is to introduce an expanding or contracting shell of ionized
gas, in which each component comes from either front or back side of the shell.
A near-systemic component may be attributed to other component, e.g., a dusty
halo without violent motion or scattering the companion nuclear light, although
we can not discriminate these ideas.

\subsubsection{The Superwind Model}

Because of the vigorous star-formation activity of the NW companion, it seems
natural to expect a superwind activity associated with the companion.
The superwind blows at later phase of the starburst evolution when the large
number of OB stars die as supernovae (SNe) and release huge kinetic energy
into circumnuclear region, and a circumnuclear bubble of shock-heated hot
ionized gas expands eventually out to halo area (see, e.g., Heckman, Armus,
\& Miley 1990 for details).
During the course of expansion, the bubble interacts with dense gas within
the galactic disk, and it preferentially elongates along the disk polar
direction where the density is smallest.
The nebula emits UV-optical emission lines (including Ly$\alpha$) by shock
heating occurring around the hot gas bubble where it interacts with ambient
cold matter.
Therefore, the superwind nebula often forms an expanding shell or bubble,
being capable of producing a pair of blue- and red-shifted nebula emissions.
Since the companion shows a ``linear'' or highly elongated disk-like structure
in continuum along NW-SE direction (Hu et al. 1996), the superwind nebula,
if any, would show elongation along NE-SW direction, being consistent with
the observation.
Superwind nebula can extend as large as a few kpc $-$ several tens kpc, and
the wind velocity can be as fast as a few $-$ several hundreds km s$^{-1}$
(e.g., Heckman et al. 1990).
Therefore, the superwind model can explain velocity structure and morphology
of the nebula in a qualitative way.
An axis of the superwind outflow is likely to be close to the sky plane, as
expected from the highly elongated appearance of the host galaxy, and this
could help to explain weak velocity shear along the nebula extension.

According to the galactic wind theory (e.g., Arimoto \& Yoshii 1987), the star
formation ceases when the galactic wind blows out.
Therefore, the youngest stellar population, which has born when the wind just
started to blow out, will contribute most to the observed UV light due to
largest UV luminosity/mass ratio, if SFR is almost constant while star-bursting.
Since the blue UV color indicates an age of the stellar population as
$\sim 10$ Myr, the wind age (or timescale of the wind propagation) is likely
to be $\sim 10$ Myr.
If this is the case, the superwind nebula could extend out to $\simeq 4$ kpc/$f$
(or $\simeq 0.6$\arcsec/$f$), assuming a constant wind speed of
$V_{\rm exp} =$ ($1/2$ of velocity separation of blue- and red-Gaussian
components at off-nuclear region)/$f \simeq 400$ km s$^{-1}$/$f$, where $f$
is a geometrical conversion factor for calculating transverse velocity from
the line-of-sight velocity.
This estimate is consistent with the observed typical nebula size
($\simeq 3$\arcsec) if $f$ is $\sim 1/5$.
Therefore, all information (stellar color, line width, and nebula size) can be
understood in a context of the superwind model, although all the estimates are
based on a simple and/or order-of-magnitude calculation.

\subsection{Origin of Diffuse Continuum Component around the Companion}

We showed that the continuum emission extends spatially along the NE-SW slit
over $\pm 3$\arcsec~from the companion nucleus.
Since the Ly$\alpha$ nebula shows more concentrated flux distribution around
the nucleus, this component should show a continuum-dominated
spectrum at $\lambda > \lambda_{\rm Ly\alpha}$, especially at SW side of
the companion (Figure 3).
Therefore, neither of scattering of the companion light, photoionization
either by the quasar nor the companion are likely as an origin of the component,
since all these models would create a Ly$\alpha$-dominated spectrum.
Therefore, the most likely origin of the component is a scattering of
the quasar light at circumgalactic matter.

The HST image also reveals a diffuse continuum structure around the companion
(Figure 1).
This component seems to show similar morphology to that of the Ly$\alpha$
nebula, i.e., the elongation is found along NE-SW direction but not
toward NW (information toward SE is lost due to bright quasar light).
Therefore, the medium for the quasar light scattering is likely to be associated
with the NW companion.
Here, a question arises as how such a component is created around the companion.
There are several possibilities:
(1) tidal structures around the companion made during merging process of
the companion, quasar, and the 2nd CO emitter,
(2) pre-galactic clumps remaining around the companion,
(3) dusty ejecta from the companion, and
(4) genuine dusty halo of the companion.
Although we could not discriminate these possibilities, the third possibility
seems more likely, because neutral matter could be transferred from dusty
nuclear region out to halo as well as the ionized gas within a context of
superwind model (Heckman et al. 2000 and references therein).
If this is the case, we can explain similar morphological properties between
Ly$\alpha$ nebula and the continuum structure.
Because there are plenty of molecular gas (and hence dusts) within the BR
1202-0725 group (at quasar and 2nd CO emitter), it seems natural to consider
that the star formation occurred in a dusty environment also in the companion.
If we assume that the dusty circumnuclear material has been expelled by
the superwind outflow, we might explain a current weak-reddening nuclear
environment of the companion.

\clearpage

\figcaption{
Slit positions are shown on a HST F814W (left) and a narrow-band Ly$\alpha$
(middle) images of Hu et al. (1996).
The HST F814W image was made by us from archival data.
A radio-continuum image at 1.35 mm (Guilloteau et al. 1999) is superimposed
on the F814W image.
All these images are shown at a same scale.
Positions of the NW companion seen in the F814W and the Ly$\alpha$ images are
marked with triangle and square on the F814W image, respectively.
Ly$\alpha$ spectrograms are shown along each slit position.
The spectrograms are smoothed with a $3\times 3$ boxcar kernel to
enhance the velocity structure at fainter and outer nebulae.
}

\figcaption{Observed spectra of quasar (top) and the NW companion (bottom).
The bottom panel contains sub-panels for close-ups of the Ly$\alpha$ profile
and the continuum spectrum, as well as the scaled sky spectrum (in red) and
the atmospheric transmission curve (in blue).
In a sub-panel of the Ly$\alpha$ profile, a fitted Gaussian emission
profile is overlaid on the observed profile (in green), as well as
a horizontal tick mark showing the instrumental spectral resolution (in red).
Expected positions for emission/absorption lines are also indicated along
the spectrum.
}

\figcaption{
Averaged spatial flux distributions are shown along NW-SE slit (upper) and
NE-SW slit (below) for Ly$\alpha$ (in black) and continuum (in blue),
separately.
In each plot, dashed blue lines show the continuum flux distribution scaled
to match that of the Ly$\alpha$ near their peaks, which make it easier
to compare the spatial flux distributions of Ly$\alpha$ and continuum.
Relative flux scales for both panels are same.
Positions of accompanying objects (2nd CO emitter and the quasar) are marked
in the upper panel.
}

\figcaption{
Ly$\alpha$ profiles are shown along each slit at 0.6\arcsec~step.
Results of the profile decomposition at 0.6\arcsec~NE nebula are overlaid on
the observed spectrum, with blue- and red-shifted Gaussian components
(in blue and red, respectively) and their sum (in green), as well as
the fitting residual (in pink) and the zero flux level (in yellow).
}

\figcaption{
Comparison of the observed continuum luminosities and synthetic models are
made.
Observed $I$ (in black) and $K$ (in pink) luminosities are shown as well as
[O {\sc ii}] emission-corrected stellar $K$ luminosity (in black).
Synthetic stellar continua of the instantaneous burst models at various ages
($5-50$ Myr) are shown for the cases of metallicities of $Z=1/5Z_{\rm \sun}$
(upper) and $Z=1/20Z_{\rm \sun}$ (lower).
All model continua are normalized at $I$ band.
}


\clearpage

\begin{references}
\reference{1}{Ajiki, M., et al. 2002, \apj, 576, L25}
\reference{1}{Arimoto, N., \& Yoshii, Y. 1987, \aap, 173, 23}
\reference{1}{Benford, D. J., Cox, P., Omont, A., Phillips, T. G., \&
McMahon, R. G. 1999, \apj, 518, L65}
\reference{1}{Carilli, C. L., Kohno, K., Kawabe, R., Ohta, K., Henkel, C.,
Menten, K. M., Yun, M. S., Petric, A., \& Tutui, Y. 2002, \aj, 123, 1838}
\reference{1}{Dawson, S., Spinrad, H., Stern, D., Dey, A., van Breugel, W.,
de Vries, W., \& Reuland, M. 2002, \apj, 570, 92}
\reference{1}{Djorgovski, S. G. 1995, in Science with the VLT, ed. J. R. Walsh
\& I. J. Danziger (Berlin: Springer), 351}
\reference{1}{Hu, E. M., McMahon, R. G., \& Egami, E. 1996, \apj, 459, L53}
\reference{1}{Hu, E. M., McMahon, R. G., \& Egami, E. 1997, in The Hubble
Space Telescope and the High Redshift Universe, ed. N. R. Tanvir,
A. Aragon-Salamanca, \& J. V. Wall (Singapore: World Scientific), 91}
\reference{1}{Fontana, A., D'Odorico, S., Giallongo, E., Cristiani, S.,
Monnet, G., \& Petitjean, P. 1998, \aj, 115, 1225}
\reference{1}{Guilloteau, S., Omont, A., Cox, P., McMahon, R. G., \&
Petitjean, P. 1999, \aap, 349, 363}
\reference{1}{Iye, M., et al. 2004, \pasj, 56, 381}
\reference{1}{Haiman, Z., Spaans, M., \& Quataert, E. 2000, \apj, 537, L5}
\reference{1}{Heckman, T. M., Armus, L., \& Miley, G. K. 1990, \apjs, 74, 833}
\reference{1}{Heckman, T. M., Lehnert, M. D., Strickland, D. K., \& Lee, A.
2000, \apjs, 129, 493}
\reference{1}{Kashikawa, N., et al. 2002, \pasj, 54, 819}
\reference{1}{Kennefick, J. D., Djorgovski, S. G., \& Meylan, G. 1996, \aj,
111, 1816}
\reference{1}{Kennicutt, R. C., Jr. 1998, \araa, 36, 189}
\reference{1}{Kodaira, K., et al. 2003, \pasj, 55, L17}
\reference{1}{Leitherer, C., Schaerer, D., Goldader, J. D., Delgado, R. M.,
Gonz\'alez, R., Robert, C., Kune, D. F., de Mello, D. F., Devost, D., \&
Heckman, T. M. 1999, \apjs, 123, 3}
\reference{1}{Mas-Hesse, J. M., Kunth, D., Tenorio-Tagle, G., Leitherer, C.,
Terlevich, R. J., \& Terlevich, E. 2003, \apj, 598, 858}
\reference{1}{Ohta, K., Yamada, T., Nakanishi, K., Kohno, K., Akiyama, M., \&
Kawabe, R. 1996, \nat, 382, 426}
\reference{1}{Ohta, K., Nakanishi, K., Akiyama, M., Yamada, T., Kohno, K.,
Kawabe, R., Kuno, N., \& Nakai, N. 1998, \pasj, 50, 303}
\reference{1}{Ohta, K., Matsumoto, T., Goto, M., Motohara, K., Taguchi, T.,
Hata, R., Yoshida, M., Iye, M., Simpson, C., \& Takta, T. 2000, \pasj, 52, 557}
\reference{1}{Omont, A., Petitjean, P., Guilloteau, S., McMahon, R. G.,
Solomon, P. M., \& P\'econtal, E. 1996, \nat, 382, 428}
\reference{1}{Pahre, M. A., \& Djorgovski, S. G. 1995, \apj, 449, L1}
\reference{1}{Petitjean, P., P\'econtal, E., Valls-Gabaud, D., \& Charlot, S.
1996, \nat, 380, 411}
\reference{1}{Rhoads, J. E., Dey, A., Malhotra, S., Stern, D., Spinrad, H.,
Jannuzi, B. T., Dawson, S., Brown, M. J. I., \& Landes, E. 2003, \aj, 125,
1006}
\reference{1}{Shull, J. M., \& McKee, C. F. 1979, \apj, 227, 131}
\reference{1}{Storrie-Lombardi, L. J., McMahon, R. G., Irwin, M. J., \&
Hazzard, C. 1996, \apj, 468, 121}
\reference{1}{Taniguchi, Y., Shioya, Y., Fujita, S. S., Nagao, T.,
Murayama, T., \& Ajiki, M. 2003, JKAS, 36, 123 (astro-ph/0306409);
Erratum, JKAS, 36, 283}
\reference{1}{Yun, M. S., Carilli, C. L., Kawabe, R., Tutsui, Y., Kohno, K.,
\& Ohta, K. 2000, \apj, 528, 171}
\end{references}
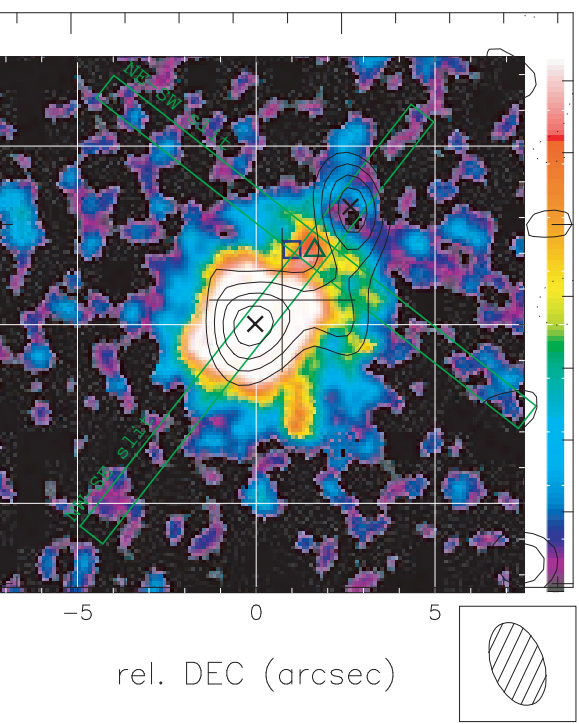
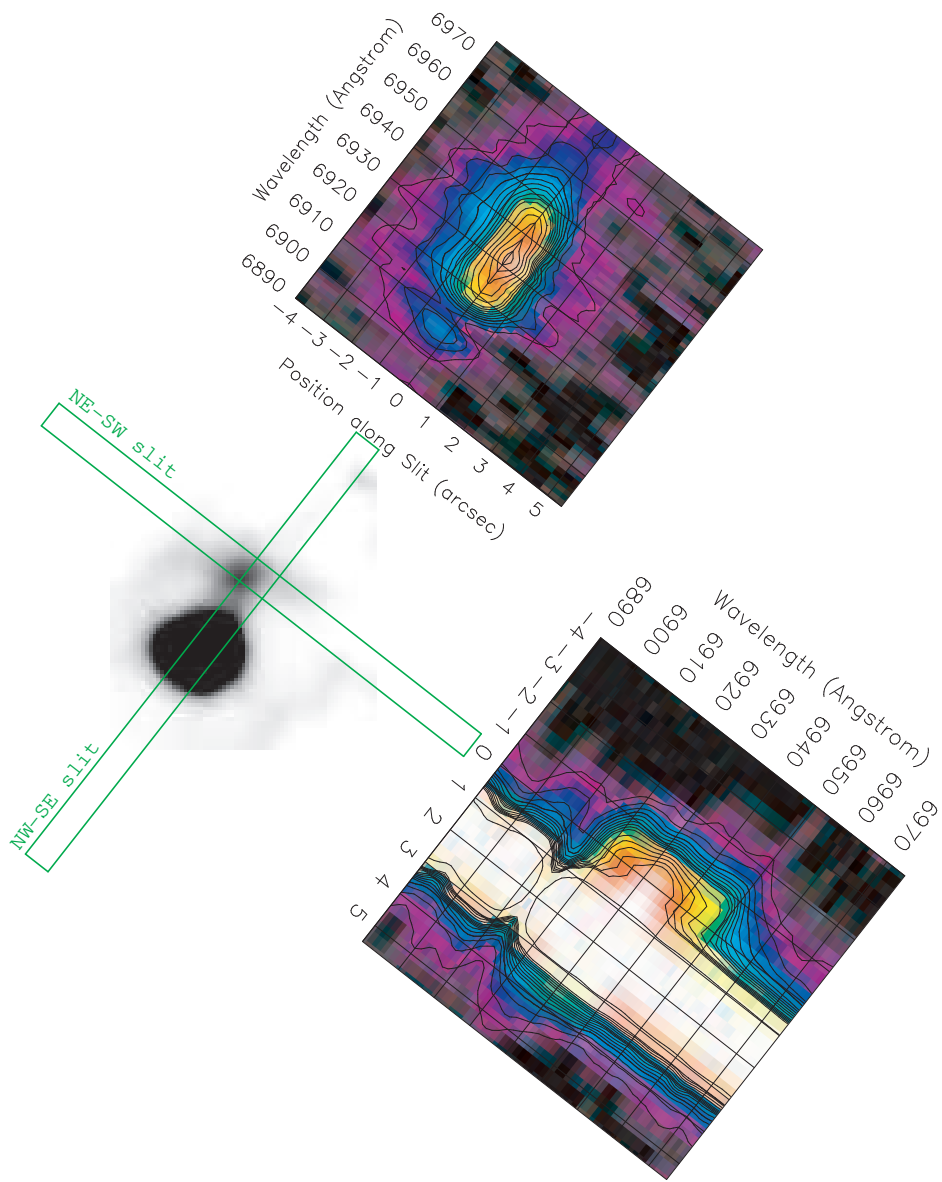
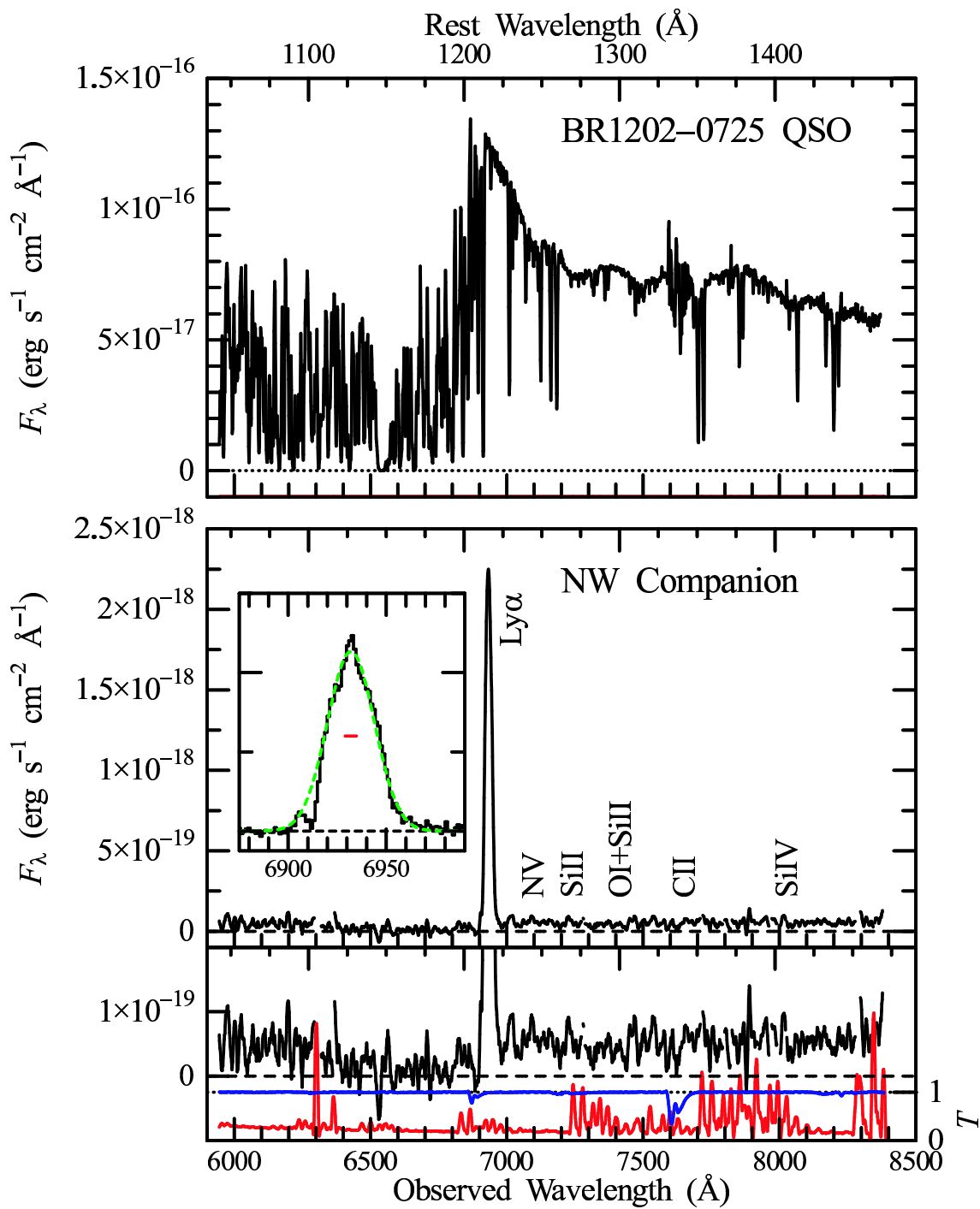
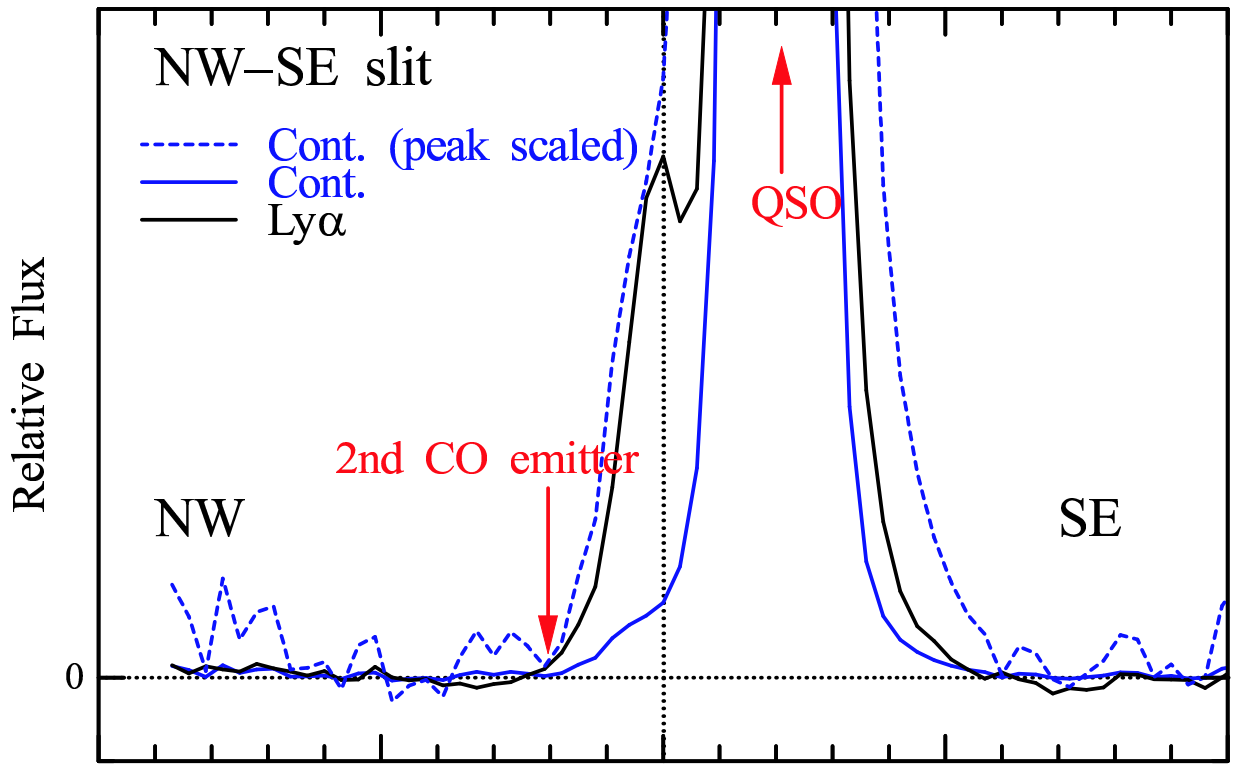
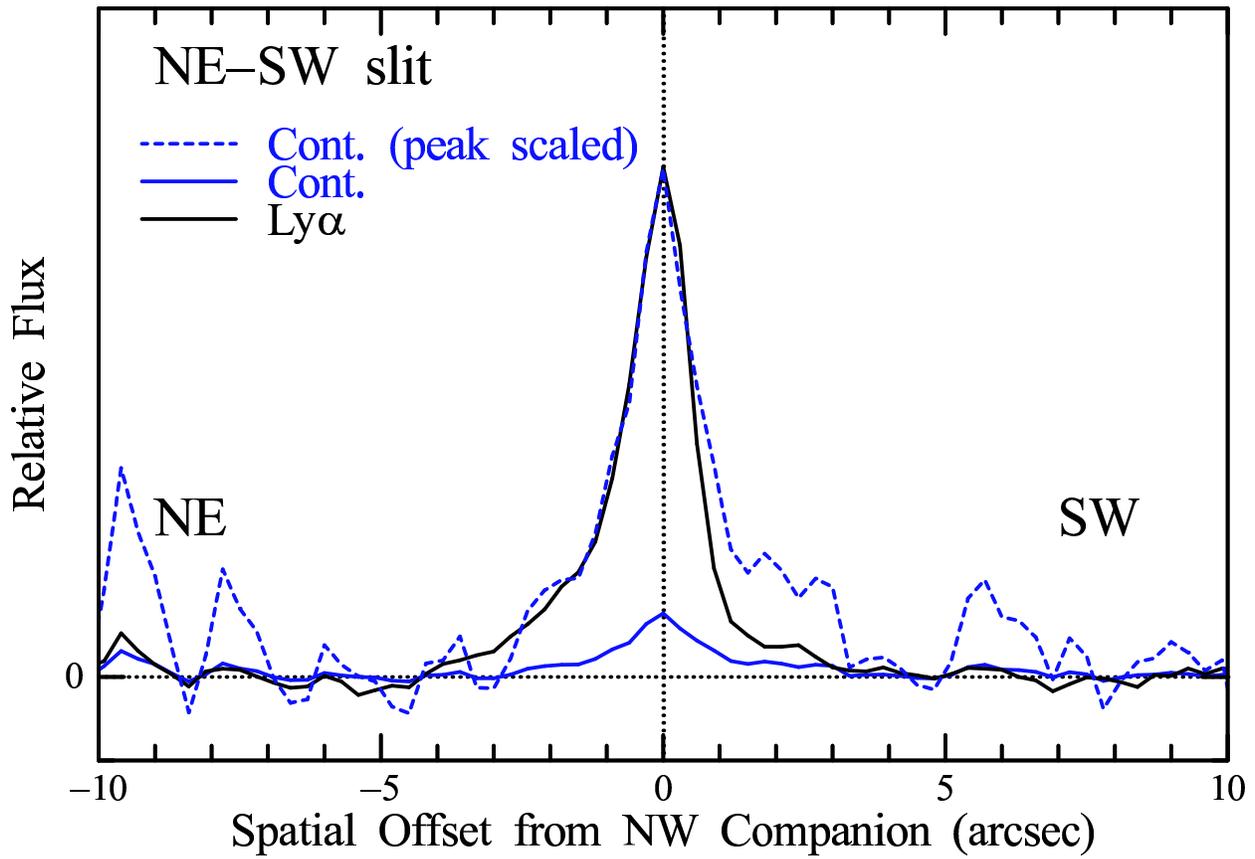
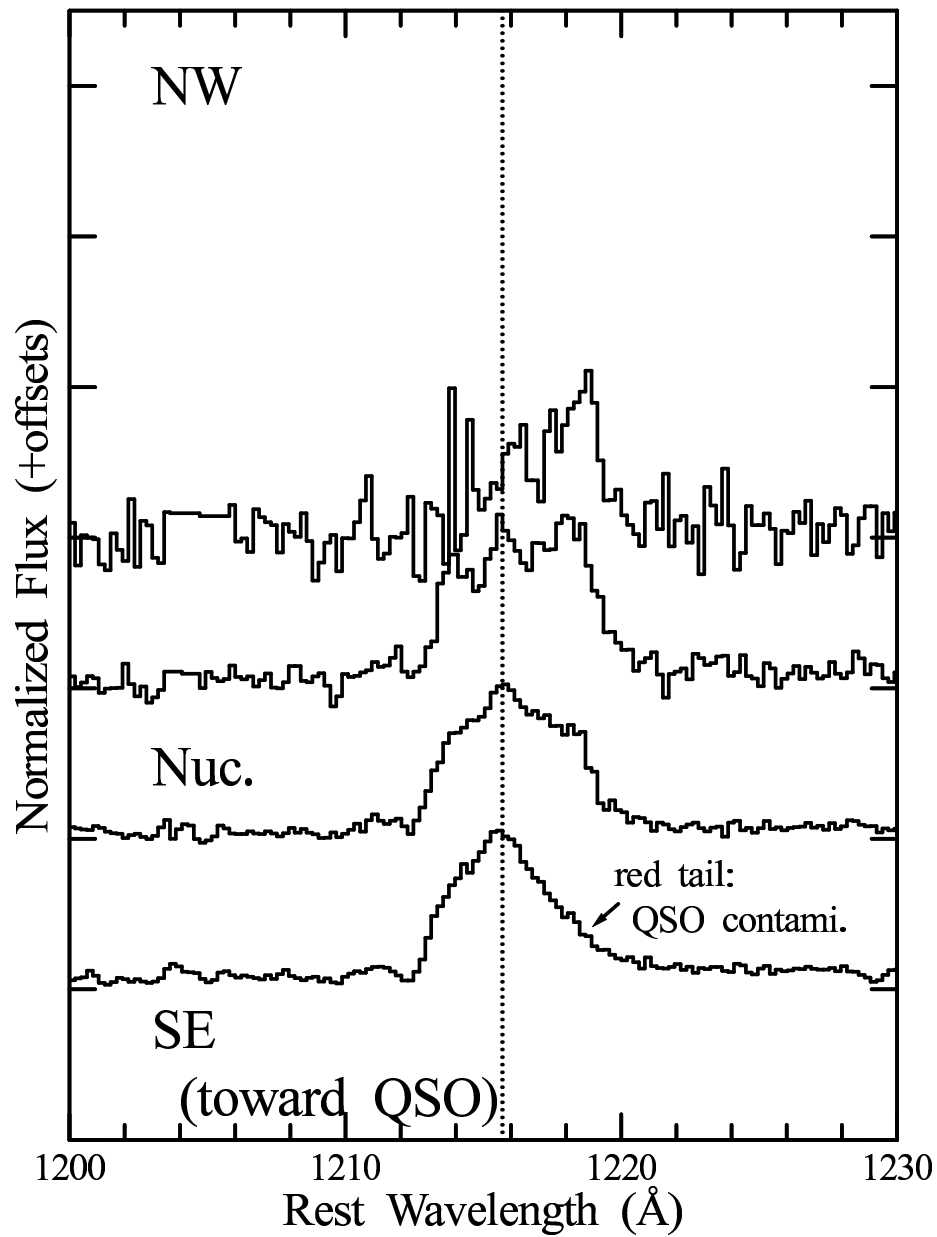
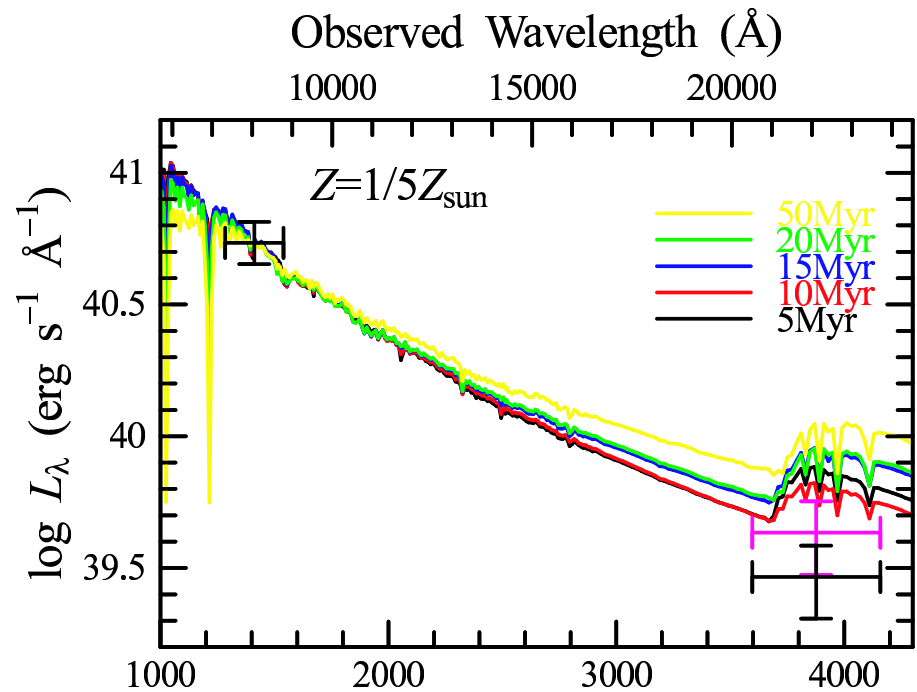
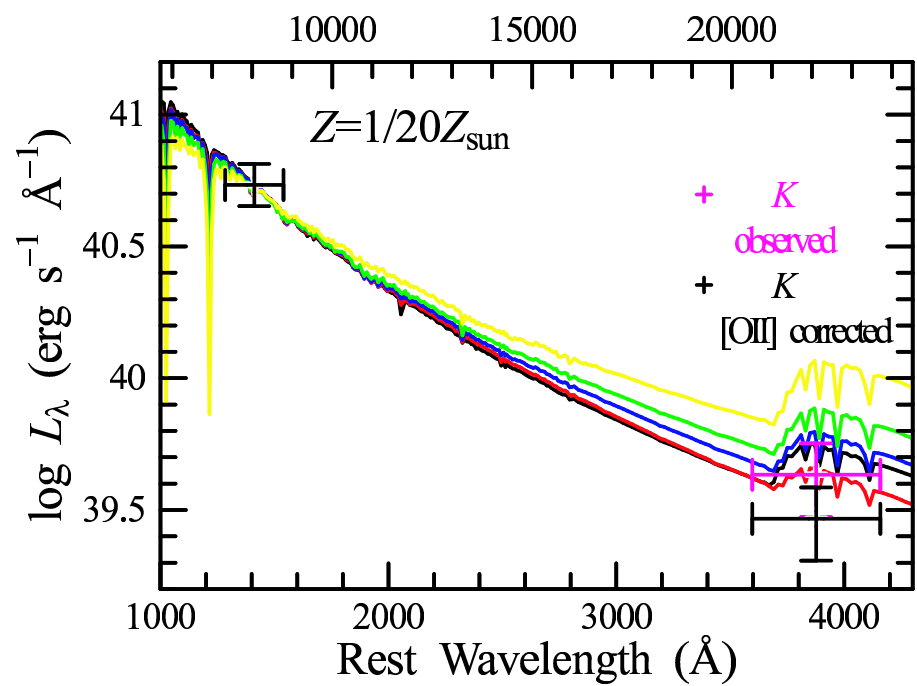
\end{document}